%% file: main.tex
\documentclass[sigconf,9pt]{acmart}

\renewcommand\footnotetextcopyrightpermission[1]{}
\usepackage{tikz}
\usepackage{braket}
\usepackage{lipsum}
\usepackage{subfig}
\usepackage{xcolor}
\usepackage{amsmath}
\usepackage{balance}
\usepackage{environ}
\usepackage{physics}
\usepackage{colortbl}
\usepackage{enumitem}
\usepackage{amsfonts}
\usepackage{booktabs}
\usepackage{graphicx}
\usepackage{textcomp}
\usepackage{algorithm}
\usepackage{algpseudocode}
\usepackage[normalem]{ulem}

\newcommand{\sol}{ReCon}

\newcommand{\genout}{$\ket{\Psi}$}

\newcommand{\fidpercentage}{33\%}

\begin{document}

\title{\sol{}: Reconfiguring Analog Rydberg Atom Quantum Computers for Quantum Generative Adversarial Networks}

\author{Nicholas S. DiBrita, Daniel Leeds, Yuqian Huo, Jason Ludmir, and Tirthak Patel}
\affiliation{
\institution{Rice University}
\city{Houston}
\state{TX}
\country{USA}
}

\begin{abstract}

Quantum computing has shown theoretical promise of speedup in several machine learning tasks, including generative tasks using generative adversarial networks (GANs). While quantum computers have been implemented with different types of technologies, recently, analog Rydberg atom quantum computers have been demonstrated to have desirable properties such as reconfigurable qubit (quantum bit) positions and multi-qubit operations. To leverage the properties of this technology, we propose \sol{}, the first work to implement quantum GANs on analog Rydberg atom quantum computers. Our evaluation using simulations and real-computer executions shows \fidpercentage{} better quality (measured using Frechet Inception Distance (FID)) in generated images than the state-of-the-art technique implemented on superconducting-qubit technology.

\end{abstract}



\maketitle

\pagestyle{plain}

\input{sections/introduction}
\input{sections/background}

\input{sections/motivation}
\input{sections/design}

\input{sections/methodology}
\input{sections/evaluation}
\input{sections/related_work}

\input{sections/conclusion}

\balance

\bibliographystyle{ACM-Reference-Format}
\bibliography{main}

\end{document}

%% file: sections/introduction.tex
\section{Introduction}
\label{sec:design}

Quantum computing has the theoretical promise of accelerating applications in multiple computing domains, including cryptography, combinatorial optimization, and machine learning~\cite{preskill2018quantum,preskill2021quantum,zeng2019learning}. One such machine learning application is generative models such as generative adversarial networks (GANs), which train a generator and a discriminator against each other to generate varied data that mimics the qualities of the input dataset~\cite{creswell2018generative}. GANs have a variety of use cases, including generating images for healthcare models, bias reduction, data augmentation, and corrective filtering~\cite{karras2018progressive, marafioti2019adversarial,blanchard2021using}. GANs have also been shown to theoretically receive quadratic speedup from a quantum computing implementation~\cite{zeng2019learning}. However, there are several challenges that need to be addressed for the theoretical promise to become a reality.

\vspace{2mm}

\noindent\textbf{The Opportunity Gap.} While GANs can theoretically benefit from quantum computing, current noisy intermediate-scale quantum (NISQ) computers suffer from a variety of challenges related to hardware errors and constraints~\cite{patel2020disq,Henriet2020-kl,patel2020ureqa}. For instance, prior quantum GANs have been implemented on superconducting-qubit technology, which has benefited from rapid development due to being built on readily-available silicon subtrates~\cite{li2020towards,arute2019quantum,smith2022scaling}. However, this technology has challenges related to short qubit coherence times, lack of operations involving more than two qubits, and hardwired qubit connectivity~\cite{burnett2019decoherence,li2020towards,patel2020experimental}. While the recently demonstrated Rydberg atom quantum computers also suffer from hardware errors, they have advantages such as longer qubit coherence times, multi-qubit operations, and reconfigurable qubit connectivity~\cite{Henriet2020-kl,Saffman2019-nd,tan2022qubit}.

\vspace{2mm}

\noindent\textit{To overcome the challenges of hardware errors and leverage the advantages of Rydberg atom technology, \sol{} develops the first solution to execute quantum GANs on Rydberg atom quantum computers.}

\vspace{2mm}

\noindent\textbf{Overview of \sol{}\footnote{\sol{} is published in the Proceedings of the International Conference on Computer-Aided Design (ICCAD), 2024.}.} As with previous quantum GAN works~\cite{dallaire2018quantum,huang2021experimental,hu2019quantum,tsang2023hybrid,silver2023mosaiq}, \sol{} takes the approach of having a quantum generator and a classical discriminator. To implement its design on NISQ computers, \sol{} first processes the input image dataset using Principal Component Analysis (PCA) for dimensionality reduction. This allows \sol{}'s generator to be implemented with only four qubits. For the generator design, \sol{} parameterizes the positions of the qubits, as well as the pulse profiles of the local (per-qubit) detuning and the global frequency (not available at a per-qubit granularity in hardware) that govern the system's Hamiltonian evolution. \sol{} leverages a selective ensemble of diverse learners with different pulse shapes to generate images of different varieties. Finally, \sol{} performs layered parameter training for parameter space reduction and improved image quality.

\vspace{2mm}

\noindent\textbf{The contributions of the work are as follows.}

\begin{itemize}[leftmargin=*]

    \item \sol{} is the first work to demonstrate quantum GANs on analog Rydberg atom quantum computers. \sol{}'s small-scale design and careful manipulation of reconfigurable features of the Rydberg atom technology make this demonstration possible.

    \vspace{1mm}
    
    \item \sol{} contributes optimizations such as noise injection at pulse starting points, parameterization of spatial and temporal features, and layered ensemble training and learner selection for improved quality and variety in generated images.

    \vspace{1mm}

    \item \sol{} performs experimental evaluation using ideal and error-prone simulation, with MNIST~\cite{mnist} and Fashion-MNIST~\cite{xiao2017/online} datasets, to achieve \fidpercentage{} lower Frechet-Inception Distance (FID)~\cite{heusel2017gans} (as measured on MNIST) than MosaiQ~\cite{silver2023mosaiq}, a state-of-the-art technique implemented on superconducting-qubit computers.

    \vspace{1mm}

    \item \sol{} is also executed on the QuEra Aquila computer~\cite{wurtz2023aquila} available via AWS Braket~\cite{gonzalez2021cloud} and shows that it generates relatively high-quality images even in the noisy real-computer setup.

    \vspace{1mm}

    \item \sol{}'s training and inference setups and models, as well as experimental evaluation results, are available for open-source usage at: \textit{\url{https://github.com/positivetechnologylab/ReCon}}.
\end{itemize}

%% file: sections/background.tex
\section{Relevant Background}
\label{sec:design}

\noindent \textbf{Quantum States and Probabilities.}
Quantum computers use the qubit as the fundamental unit of information. We represent the state of a qubit as a \textit{ket}, written as $\ket{\psi}$, existing in a 2-dimensional complex vector space (specifically a Hilbert space) with basis states $\ket{0}$ and $\ket{1}$. The state of the qubit can thus be written generally as $|\psi\rangle = \alpha|0\rangle + \beta|1\rangle$, where the coefficients $\alpha$ and $\beta$ are complex numbers satisfying $|\alpha|^2 + |\beta|^2 = 1$. These coefficients are important in that they encode the probability of measuring either 0 or 1, the probability of the former being given by $\abs{\alpha}^2$ and the latter by $\abs{\beta}^2$.

This representation generalizes to systems having higher qubit counts. We combine the individual Hilbert spaces of qubits via the tensor product operation.
For two qubits, this results in a combined 4-dimensional space that is spanned by basis vectors $|00\rangle$, $|01\rangle$, $|10\rangle$, and $|11\rangle$. We can extend this idea to more than two qubits: for example, a 3-qubit system will have an 8-dimensional Hilbert space spanned by $|000\rangle$, $|001\rangle$, $|010\rangle$, $|011\rangle$, $|100\rangle$, $|101\rangle$, $|110\rangle$, and $|111\rangle$. In general, $n$ qubits will have a combined state $\ket{\Psi} = \sum_{k=0}^{k=2^n-1}\alpha_k\ket{k}$, such that $\sum_{k=0}^{k=2^n-1}\abs{\alpha_k}^2 = 1$. The probability of observing the $k^{th}$ state is $\abs{\alpha_k}^2$. The Hilbert space size for an $n-$qubit system is $2^n$.

Measurement of quantum states is an operation that projects a system of qubits onto a single basis state. When the system's state is measured, the state collapses into one of the basis states with a probability equal to the absolute square of the amplitude associated with that basis. Following measurement, the qubit system irreversibly becomes the observed basis state.  Thus, given a vector in a Hilbert space (which represents a qubit system), each of the basis states is assigned a probability amplitude such that the square of the absolute value of the probability amplitude represents the probability of that state occurring, and the sum of the squared absolute values of all probability amplitudes for all basis states is 1.

\begin{figure}
    \centering
    \subfloat[Analog Quantum Model]{\includegraphics[scale=0.355]{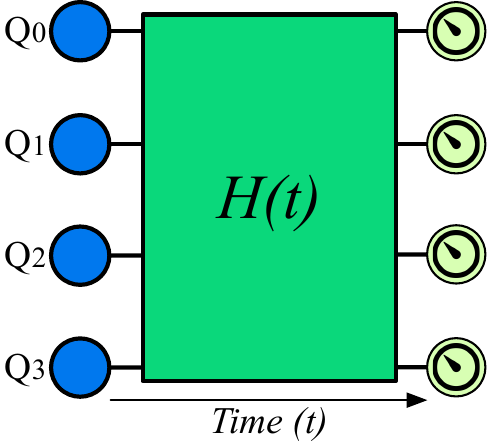}}
    \hfill
    \subfloat[Position-Dependent Interaction Strength]{\includegraphics[scale=0.287]{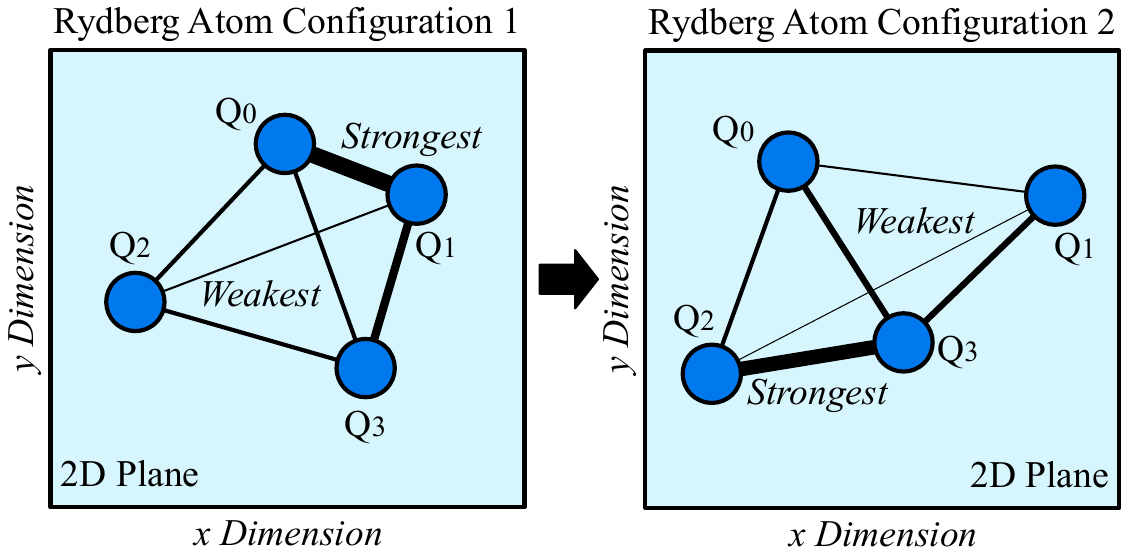}}
    \vspace{0.5mm}
    \hrule
    \vspace{-3.5mm}
    \caption{(a) The analog quantum model involves the evolution of quantum register under some (time-dependent) Hamiltonian $(H)$ followed by qubit measurement at the end (circular meters). (b) Atoms can be reconfigured into different positions that affect their interaction strength.}
    \label{fig:positions}
    \vspace{-4mm}
\end{figure}

Logical operations on qubits consist of unitary operations that modify the amplitude and phase of the coefficients of an $n$-qubit system. In a digital quantum computer, these operations are represented as discrete gates which perform a well-defined transformation. Sequences of quantum gates are arranged to form quantum circuits, which realize the functionality of quantum algorithms. 

\vspace{2mm}

\noindent \textbf{Rydberg Atom Quantum Computers.} In a Rydberg atom computer, qubits are stored as the electronic energy levels of neutral atoms~\cite{Henriet2020-kl, wurtz2023aquila}. Generally, these computers use a low energy ground state, $\ket{g}$, and a highly excited state $\ket{r}$, referred to as a Rydberg state. One possible scheme uses these two states as our qubit states upon defining $\ket{0} = \ket{g}$ and $\ket{1} = \ket{r}$. We will use these notations interchangeably. We also use the terms ``qubits'' and ``atoms'' interchangeably, as they represent the same entity in this context.

Rydberg atom quantum computers utilize the distinct properties of these Rydberg states, which are obtained by exciting atoms through precision laser pulses. In these excited Rydberg states, atoms exist with large atomic radii and enhanced dipole moments. Such attributes enable long-range interactions between atoms, which can be used for entanglement and quantum computing operations~\cite{Henriet2020-kl}. These Rydberg interactions increase the energy cost of exciting nearby atoms to their respective Rydberg states. If this increase in energy is large enough, atoms are prevented from reaching the excited state. This is called the Rydberg blockade effect, and it is essential for creating multi-qubit operations~\cite{patel2022geyser}. 

\vspace{2mm}

\begin{figure}
    \centering
    \includegraphics[scale=0.30]{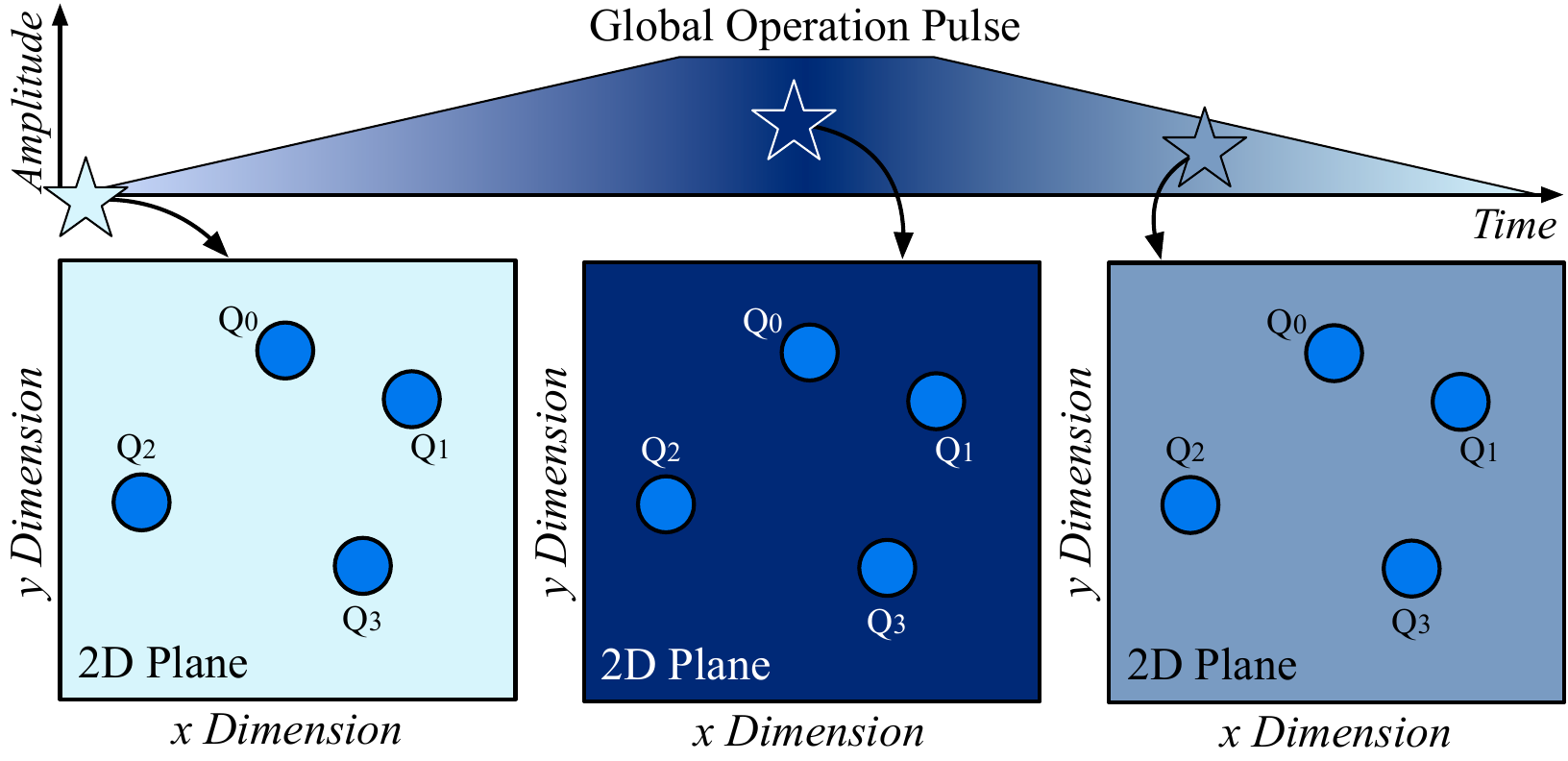}
    \vspace{0.5mm}
    \hrule
    \vspace{-3.5mm}
    \caption{Rydberg atom quantum computers can apply global operational pulses that affect all qubits simultaneously.}
    \label{fig:global}
    \vspace{-4mm}
\end{figure}

\vspace{1mm}

\noindent\textbf{Analog Quantum Computing.} Analog quantum computing utilizes the continuous time evolution of quantum systems to perform computations. The core of analog quantum computation is the \textbf{Hamiltonian} $H(t)$, which is an operator that describes the total energy of a system. The Hamiltonian plays a central role in quantum theory, as it governs the time evolution of a quantum system according to Schrödinger's equation:
\begin{equation}
i\hbar \frac{d}{dt} |\Psi(t)\rangle = H(t) |\Psi(t)\rangle
\end{equation}

Here, $|\Psi(t)\rangle$ represents the state of the quantum system at time $t$, and $\hbar$ is the reduced Planck's constant.
Unlike digital quantum computers, which define program execution in terms of discrete quantum gates, analog systems execute quantum programs by manipulating the system Hamiltonian as the state evolves in time, as shown in Fig.~\ref{fig:positions}(a). In general, Hamiltonians will have different controllable parameters, depending explicitly on the type of quantum computer used. In Rydberg atom computers, these parameters are related to the atoms' interaction with the laser and the inter-atomic spacing that controls the Rydberg interaction strength.  

\vspace{2mm}

\noindent\textbf{Advantages of Rydberg Atom Quantum Computers.}
Highly tunable Rydberg atom quantum computers provide the flexibility to finely tune the interactions between qubits to achieve optimized quantum operations. The Rydberg interaction depends inversely on the separation between atoms, so as atoms are brought closer together, they interact more strongly with one another. As demonstrated in the first configuration topology in Fig.~\ref{fig:positions}(b), Q0 and Q1 have the strongest dipole-dipole interaction, while Q1 and Q2 have the weakest. This allows the computer to strongly couple Q0 and Q1 so that their states may be entangled. The exact positioning will have important ramifications for how correlations form between qubits in different parts of the computer.

One additional feature of Rydberg atom quantum computers is the ability to enact simultaneous changes across all qubits using global operational pulses. As illustrated in Fig.~\ref{fig:global}, laser pulses can be finely tuned to impact all qubits simultaneously. \sol{} takes advantage of this drive computation and introduces variations in the process of noise generation for the image generator. 

\vspace{2mm}

\noindent\textbf{Error Effects on Rydberg Atom Quantum Computers.} Like other quantum computers, Rydberg atom computers suffer from the effects of hardware errors, which impact the execution of programs. One of the most significant sources of error comes from the decay of atoms' states. Over time, atoms in the Rydberg computer will incoherently decay into either their ground state or some intermediate state. For the Aquila computer, the Rydberg state atoms have a decoherence time of around $5.8\mu s$~\cite{wurtz2023aquila}.

Another major source of error for Rydberg atom computers comes from the measurement of qubit states. In order to measure atom states, optical traps are deactivated before operations occur and reactivated afterward. If an atom is still in its ground state, it will be re-trapped successfully. Conversely, if it is in its Rydberg state, it will be repelled. This means that when the final imaging of the atom array is done if an atom is present in a trap, this implies it is in the ground state, which in turn means the atom is in the $\ket{0}$ state; if an atom is absent from a trap, it is measured in the $\ket{1}$ state. Errors occur when a Rydberg atom is unintentionally re-trapped because its state has decayed (e.g., a $\ket{1}$-state qubit is measured as $\ket{0}$) or when an atom in the ground state is not successfully re-trapped after the optical traps are reactivated (e.g., a $\ket{0}$-state qubit is measured as $\ket{1}$).

\vspace{2mm}

\begin{figure}
    \centering
    \includegraphics[scale=0.36]{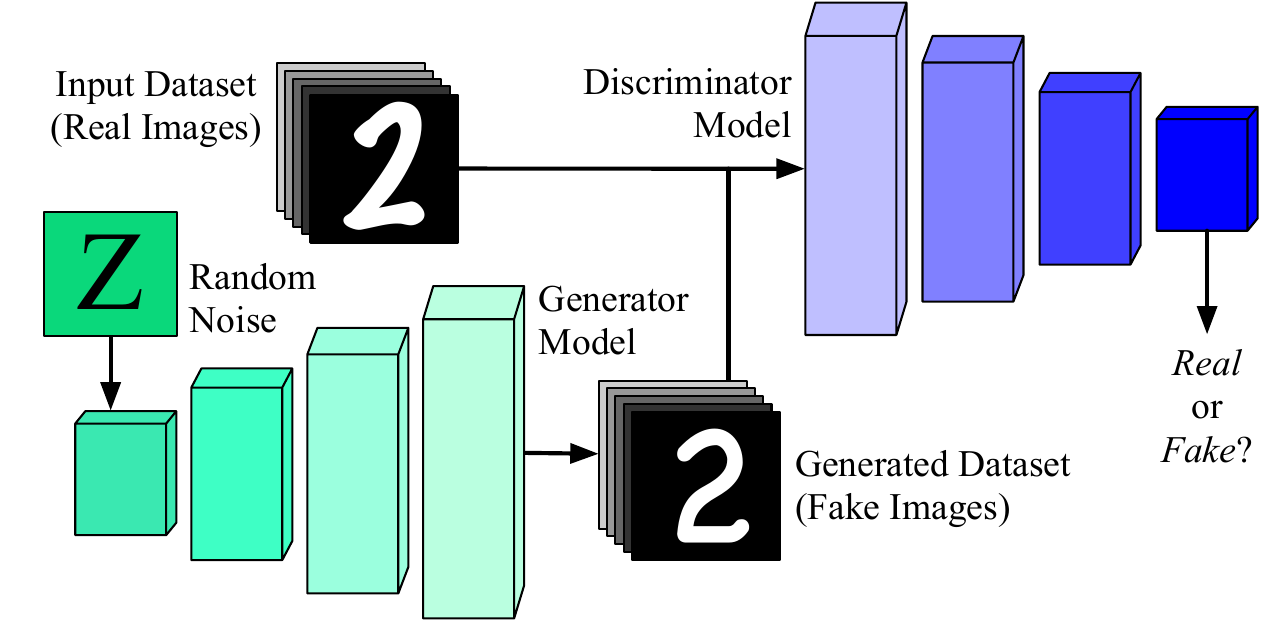}
    \vspace{0.5mm}
    \hrule
    \vspace{-3.5mm}
    \caption{Training of a classical GAN. Once trained, the generator is used to run inference tasks to generate images.}
    \label{fig:gan}
    \vspace{-4mm}
\end{figure}

\noindent\textbf{Generative Adversarial Networks(GANs).}
GANs consist of two networks: a generator and a discriminator (Fig.~\ref{fig:gan}). In a conventional image generation setting, the generator is intended to learn such that it produces synthetic images that are indistinguishable from real images. Ideally, the generator will produce a variety of high-quality images. To introduce this variability, the generator takes in seed noise, which is typically drawn from some prior distribution. The generator then maps this seed noise $Z$ to an output image $F$ by way of parameterized functions/layers:
\begin{equation}
    \label{eq:generatorfunc}
    F(Z ; \vec{\theta}) = (f_0, f_1, ..., f_k)
\end{equation}

Here, the $f_i's$ are output features (pixels in the case of image generation). The parameters $\vec{\theta}$ are then optimized so that the generator produces high-quality images across a variety of input seed noise. This is done through adversarial training against the discriminator network, which is tasked with identifying which images are real and which images are synthetic.
As the discriminator network improves in being able to distinguish between real and synthetic input images, the generator network parameters are updated by minimizing the probability that its output belongs to the synthetic class, as assigned by the discriminator network. The learning process of a GAN can thus be summarized as a back-and-forth competition between the two networks. A successfully developed generator network should be able to produce diverse, realistic images.

%% file: sections/motivation.tex
\section{Motivation for \sol{}}
\label{sec:design} 

To the best of our knowledge, we have implemented the first GANs using Rydberg atom technology. Existing quantum GANs, including MosaiQ~\cite{silver2023mosaiq}, implement digital algorithms targeting hardware like superconducting qubits. This allows them to make use of the variational quantum algorithm (VQA) formalism, which has shown promise for different kinds of optimization and machine learning tasks~\cite{cerezo2021variational,chen2020variational}. A VQA-based generator would have a predetermined digital quantum circuit, referred to as an ansatz, consisting of layers of parameterized gates. These parameters can then be optimized using classical optimization methods like gradient descent, making them an attractive choice for a variety of problems~\cite{cerezo2021variational}. 

Physically-realizable VQA circuits are limited by the inter-qubit connectivity of common hardware platforms. Solutions like MosaiQ are limited to nearest-neighbor 2-qubit logical operations, which limit the correlations that can be established between qubits and, thus, within the generator's output. This can be overcome by applying many layers of 2-qubit gates so that a given qubit can influence distant qubits through intermediary qubits; however, this increases the number of gates required and, thus, increases the errors resulting from hardware noise~\cite{silver2023mosaiq,silver2022quilt,patel2022quest}. 

\sol{} utilizes the unique properties of Rydberg atoms to overcome these limitations. In particular, it makes use of reconfigurable geometry to customize qubit connectivity uniquely. This connectivity is manifest in the strength of the Rydberg interactions discussed above. \sol{} tunes the positions of the qubits to affect their connectivity, along with the driving laser parameters, to optimize the computer's Hamiltonian evolution for image generation. 

%% file: sections/design.tex
\section{Design and Implementation}
\label{sec:design}

\begin{figure}
    \centering
    \includegraphics[scale=0.42]{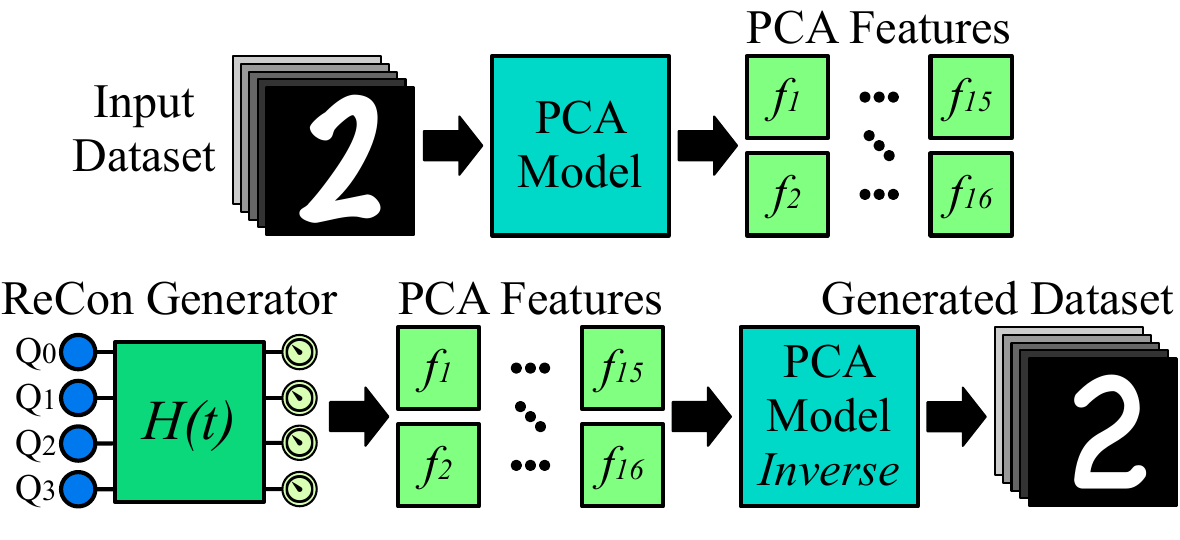}
    \vspace{-1mm}
    \hrule
    \vspace{-3.5mm}
    \caption{\sol{} uses PCA and inverse PCA in its design.}
    \label{fig:pca}
    \vspace{-3.5mm}
\end{figure}

In this section, we go through the \sol{} procedure step-by-step. 

\vspace{2mm}

\noindent\textbf{Principal Component Analysis (PCA).} \sol{} first processes the training data using Principal Component Analysis. Principal Component Analysis (PCA) is a technique that is used to reduce the dimensionality of data while retaining most of its information. This is accomplished by computing the eigenvectors of the covariance matrix of the dataset and then projecting the dataset onto these eigenvectors. Each of the eigenvectors is a principal component, which together form a new basis that simplifies the complexity of the original dataset~\cite{pearson1901pca}. We reduce the dimensionality from $28\cross28$ to $2^n$, the size of the Hilbert space for $n$ qubits. We reduce to this size so that we can make use of an amplitude encoding scheme, where the probability of measuring one of the basis states will encode one dimension of the generator's output and, thus, one of the PCA feature weights. We use $n=4$ for all results in this paper, as this system size is small enough for us to simulate the time dynamics on classical computers quickly. This allows us to train the generator practically in simulation.

\sol{}'s generator then learns to produce PCA feature weights that resemble PCA feature weights from a certain class of the training data, as shown in Fig.~\ref{fig:pca}. Likewise, the discriminator will attempt to discern between PCA feature weights produced by the generator and those obtained from the images in the training dataset. To generate digits from multiple classes, we independently train multiple generators that are each responsible for a single class. The resulting PCA transform is retained so that we can use the inverse transform to produce images from the generator output.

\begin{figure}
    \centering
    \includegraphics[scale=0.3]{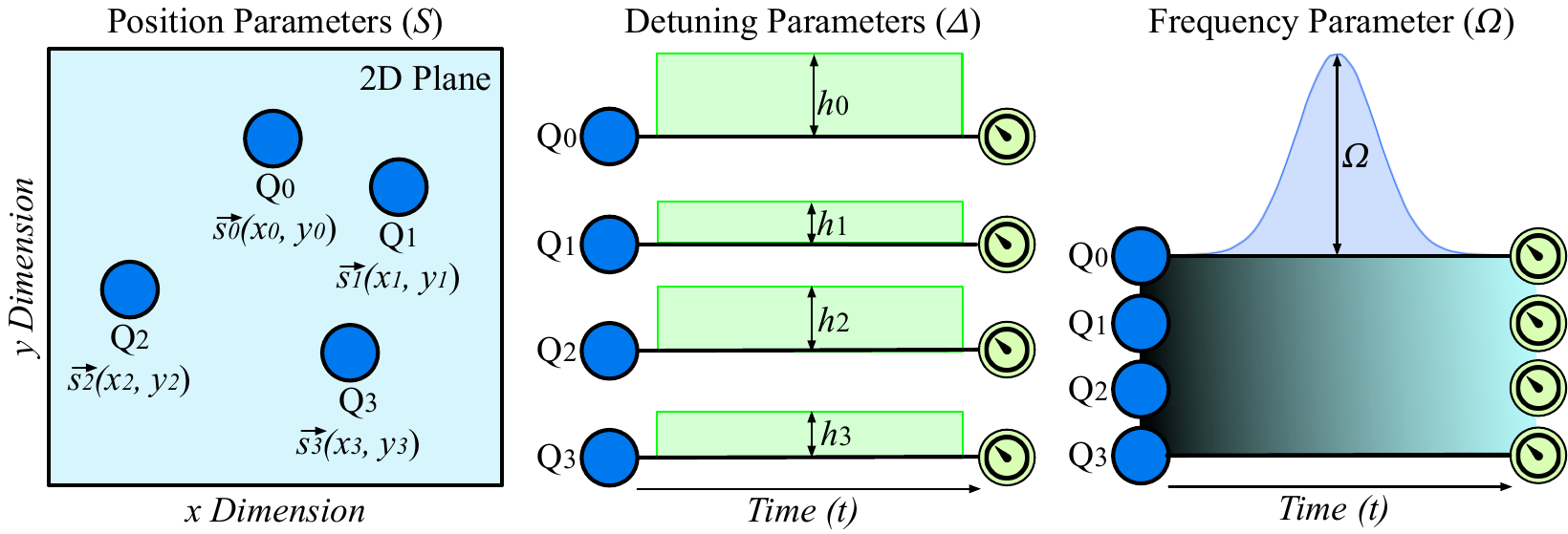}
    \vspace{0.5mm}
    \hrule
    \vspace{-3.5mm}
    \caption{\sol{} parameterizes the qubit positions (spatial), as well as the amplitudes of the local detuning, constant global detuning (not shown), and global frequency of the Hamiltonian pulses (temporal).}
    \label{fig:params}
    \vspace{-4mm}
\end{figure}

\begin{figure}
    \centering
    \includegraphics[scale=0.32]{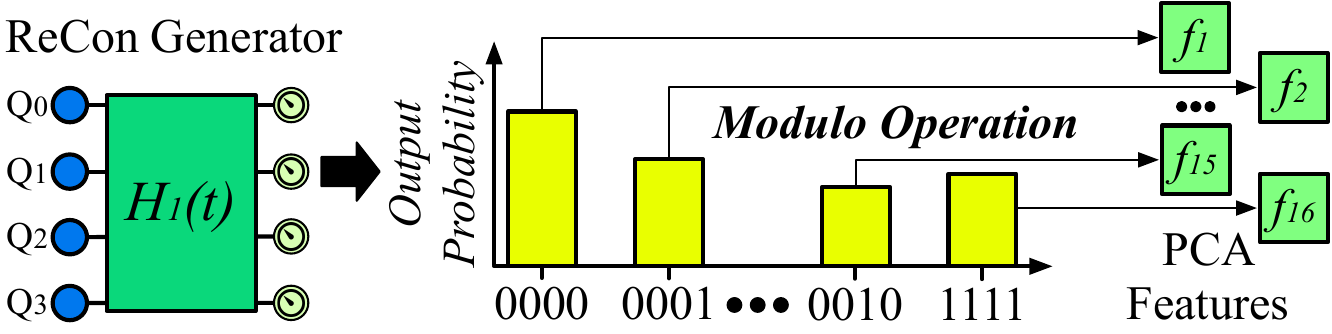}
    \vspace{0.5mm}
    \hrule
    \vspace{-3.5mm}
    \caption{\sol{} converts output probabilities of the qubit measurements into features by applying modulo operations.}
    \label{fig:mod}
    \vspace{-4mm}
\end{figure}

\begin{figure*}
    \centering
    \includegraphics[scale=0.355]{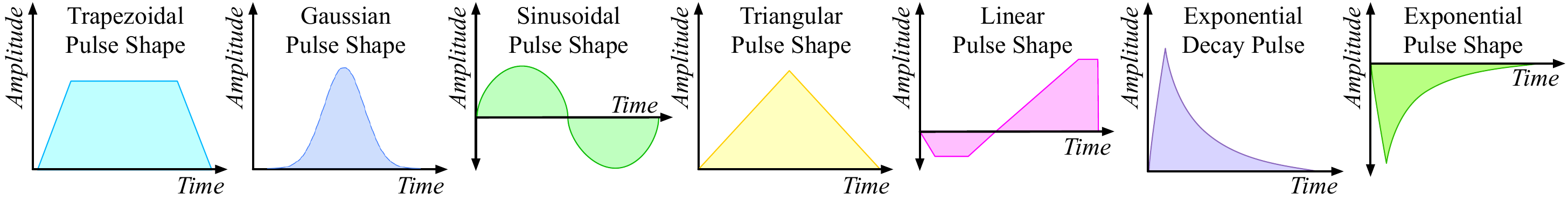}
    \vspace{0.5mm}
    \hrule
    \vspace{-3.5mm}
    \caption{\sol{} leverages pulses of different shapes to create a diverse ensemble of learners to generate a variety of images}
    \label{fig:shapes}
    \vspace{-4mm}
\end{figure*}

\vspace{2mm}

\noindent\textbf{\sol{}'s Generator Design.} \sol's generator is designed to utilize the analog capabilities of current Rydberg atom computers. The atoms are initialized in the $\ket{0000}=\ket{gggg}$ state at specified positions $S = {\Vec{s}_1,\Vec{s}_2,\Vec{s}_3,\Vec{s}_4}$ (Fig.~\ref{fig:params}). Laser pulses, along with the Rydberg interactions, then drive the system to some desired final state:
\[
\ket{\Psi} = \textstyle\sum_{i=0}^{2^n-1} \alpha_i \ket{i}
\]
This operation is the analog computation step, and following it, like prior work~\cite{hu2023tackling}, the generator output is now encoded into the probabilities of \genout, $p_i = \abs{\alpha_i}^2 = \abs{\braket{i}{\Psi}}^2$. However, naively encoding the generator output this way means that the generator outputs must all collectively sum to 1, which, in general, is not a realistic constraint. To solve this, we introduce a modulo operation (Fig.~\ref{fig:mod}):
\[
f_i = p_i \ \text{mod} \ \frac{1}{2^n}
\]
Now, each generator output $f_i$ can be any value on the interval $(0, \frac{1}{2^n}]$, regardless of the other outputs. While training, we scale the training data PCA features so they are also in this allowed range. Following training, we rescale all generated PCA features before applying the PCA inverse transform.

\vspace{2mm}

\noindent\textbf{Generator Hamiltonian Parameters.} In Rydberg atom quantum computers, laser pulses are used to manipulate the Hamiltonian, with users being able to control the pulses via three parameters: the \textbf{Rabi frequency ($\Omega$)}, \textbf{the detuning $(\Delta)$}, and the \textbf{phase ($\phi$)}. The Rabi frequency is a measure of the intensity of the laser pulse that drives the transitions between the ground state and the excited Rydberg states of qubits. This parameter controls the rate at which qubits oscillate between these two states. The detuning, $(\Delta)$, refers to the difference between the frequency of the laser and the natural frequency of the atomic transition. Finally, the relative phase $\phi$ controls the direction in which the pulse rotates each qubit's state (e.g., in the X or Y directions). By controlling these parameters over a length of time, user-defined evolutions in the Hamiltonian can be performed~\cite{Henriet2020-kl,dig_an_qc,wurtz2023aquila}. 

The globally addressed Rydberg Hamiltonian that describes the dynamics of the Rydberg quantum computer is:
\begin{equation}
\begin{split}
H(t) = \frac{\Omega(t)}{2} \sum_i \left( e^{i\phi(t)} |g\rangle_i \langle r|_i + e^{-i\phi(t)} |r\rangle_i \langle g|_i \right) \\
- \Delta_{global}(t) \sum_i \hat{n}_i + \sum_{i < j} \frac{C_6}{|\vec{s}_i - \vec{s}_j|^6} \hat{n}_i \hat{n}_j    
\end{split}
\end{equation}

In the above, $t$ represents time, and $\Omega(t)$, $\Delta_{global}(t)$, and $\phi(t)$, represent the \textit{globally addressed} Rabi frequency, detuning, and relative phase at time $t$. Here, global addressing refers to the fact that these Hamiltonian parameters are the same across all atoms.  $|g\rangle_i$ and  $|r\rangle_i$ are the ground and excited Rydberg states of atom $i$ respectively~(~$\bra{r} = \ket{r}^\dagger$, where $\dagger$ represents the Hermitian conjugate). $\hat{n}_i$~counts the number of Rydberg excitations for atom $i$ and is equivalent to $|r\rangle_i \langle r|_i$. $\frac{C_6}{|\vec{s}_i - \vec{s}_j|^6}$ represents the Rydberg interaction potential, with $\vec{s}_i$  being the location of the atom $i$ in the grid and the constant $C_6 = 5,420,503$ MHz $\mu m^6$.

Recently, it has become possible to implement additional \textit{locally addressed} detuning on the commercially available Aquila device. This adds the following local term to the above Hamiltonian:
\begin{equation}
    H_{local}(t) = -\sum_i \Delta_{local}(t) h_i \hat{n}_i
\end{equation}

Here, $\Delta_{local}(t)$ is the local detuning at time $t$, which is the same for all atoms, and $h_i$ is a site-dependent coupling to the detuning. This allows for a finer control over individual atoms, compared to the global terms above.

To summarize, the user-specified generator inputs are as follows:
\begin{enumerate}
    \item the position of each atom $\Vec{s}_i = (x_i, y_i)$
    \item the global Rabi frequency pulse $\Omega(t)$
    \item the global detuning pulse $\Delta_{global}(t)$
    \item the global phase $\phi(t)$
    \item the local detuning pulse $\Delta_{local}(t)$
    \item the local detuning coupling strength for each atom $h_i$
\end{enumerate}

The parameters of \sol{}'s generator are thus the laser pulse sequence, atom positions, and site-dependent local detuning coupling. The atomic positions and detuning coupling are both fixed at the beginning of the analog computation. The pulses are each specified by a time series. To incorporate the generator's input seed noise, we design pulses that depend non-trivially on some scalar input noise and a set of trainable parameters.

\sol{} experiments with different pulse shapes for the Rabi frequency and local detuning (Fig.~\ref{fig:params}). We fix $\phi = 0$, as it does not directly impact the amplitudes used in the generator output. We set $\Delta_{global}(t)$ to be constant in time but still have a trainable value. By doing this, it acts as a trainable ``offset'' for the local detuning, which otherwise would be constrained to a comparatively small range of possible values. For all pulse shapes, we allow the system to evolve for 1$\mu s$, as this gives a reasonable balance between allowing entanglement to build in the system and avoiding errors~\cite{lu2024digitalanalog}. 

\begin{figure}
    \centering
    \includegraphics[scale=0.37]{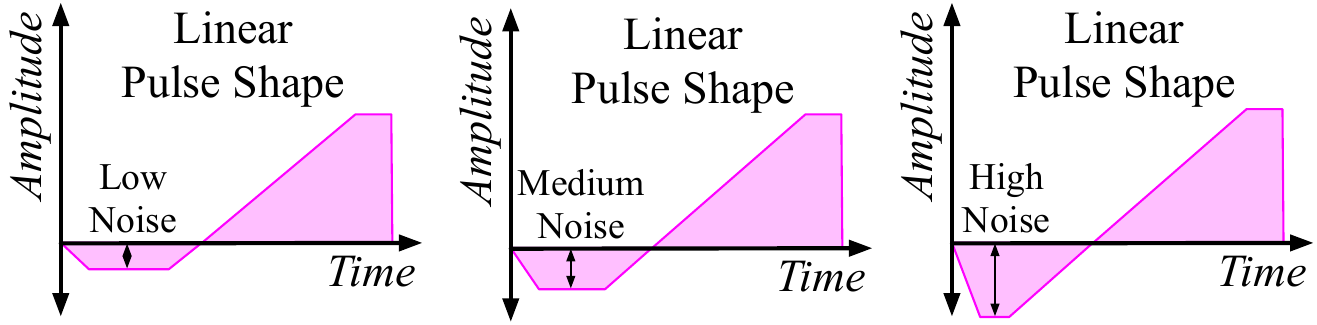}
    \vspace{0.5mm}
    \hrule
    \vspace{-3.5mm}
    \caption{\sol{} inserts noise by altering pulse initializations.}
    \label{fig:noiseinitialization}
    \vspace{-4mm}
\end{figure}

\vspace{2mm}

\noindent\textbf{Pulse Parameterization.} We design \sol{} to use a selection of simple functions as possible pulse shapes, shown in Fig.~\ref{fig:shapes}. Doing so allows \sol{} to explore a variety of different Hamiltonian evolutions without having to rely on a complex parameterization scheme. Each pulse shape depends on a single scalar parameter, which is trainable, and the input noise is used as a seed for image generation. As a concrete example, consider the linear pulse shape of Fig.~\ref{fig:noiseinitialization}. One possible parameterization is to set the initial value of the pulse based on the seed noise and then the final value based on a single scalar pulse parameter, with intermediate values determined by linear interpolation. The goal of training is then to find a suitable pulse parameter that allows for a variety of high-quality images to be made from different initial seeds. 

While there are many possible pulse shapes, there are a few constraints imposed by the specifics of the hardware implementation. In particular, $\Omega(t)$ and $\Delta_{local}(t)$ must start and end at 0. $\Omega(t)$ is required to be nonnegative, while $\Delta_{local}(t)$ must be nonpositive. The exhaustive list is available in the Amazon Braket documentation~\cite{aws}. We design pulses that obey these restrictions to be compatible with the real-hardware experiments.

\begin{figure}
    \centering
    \includegraphics[scale=0.34]{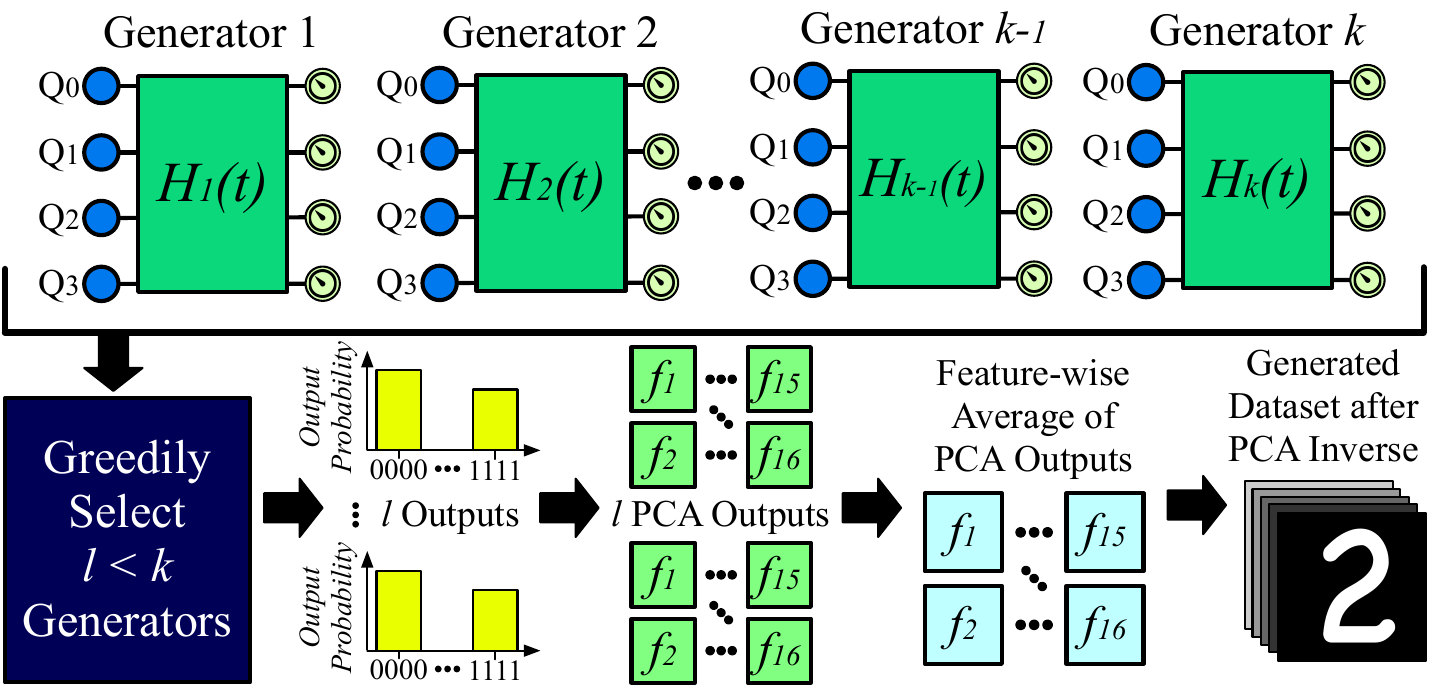}
    \vspace{0.5mm}
    \hrule
    \vspace{-3.5mm}
    \caption{\sol{} greedily selects ensemble members for its generator and averages their output for PCA features.}
    \label{fig:ensemble}
    \vspace{-3mm}
\end{figure}



\begin{algorithm}[t]
\caption{Greedy Ensemble Selection}
\begin{flushleft}
\textbf{Inputs:} $ValData$ (validation set of real data), $bestLearner$ (best learner measured against ValData)
\end{flushleft}
\hrule
\label{alg:greedy}
\begin{algorithmic}
\State $Ensemble \gets \{bestLearner\}$
\State $ValFID \gets $ FID of $bestLearner$
\State $ Searching \gets True $
    \While{$ Searching $}
        \State $nextLearner \gets nothing$
        \For{$\text{each learner } L \notin Ensemble$}
        \State $PCAbatch \gets $ Avg. of PCA weights from $L \cup Ensemble$
        \State $Images \gets $ Images from inv. PCA transform of $PCAbatch$
        \State $TrialFID \gets $ FID score of $Images$ compared to $ValData$
        \If{$TrialFID < ValFID $}
            \State $nextLearner \gets L$
            \State $ValFID \gets TrialFID$
        \EndIf{}
        \EndFor{}
        \If{$nextLearner = nothing$}
        \State $Searching \gets False$
        \Else{} 
            \State $Ensemble \gets Ensemble \cup \{nextLearner\}$
        \EndIf{}
        
    \EndWhile{}\\
\Return $Ensemble$
\end{algorithmic}
\end{algorithm}

\vspace{2mm}

\noindent\textbf{Pulse Ensembles.} One additional advantage of having multiple pulse shapes is that it naturally gives a way to implement ensemble learning. To do so, we train a variety of different learners, each of which is given a pulse for the Rabi frequency and a pulse for the local detuning. The learner is then adversarially trained against a discriminator network, and the resulting generator parameters are saved. We can then perform ensemble inference by averaging the output of each ensemble learner (member) (Fig.~\ref{fig:ensemble}).

To form an ensemble, \sol{} uses greedy forward selection. We use a validation set to compute the \textbf{Fréchet Inception Distance (FID)}~\cite{heusel2017gans} for a batch of generated images from each learner. The FID measures the similarity of two distributions (the distributions of input and generated images), with higher scores indicating more distance and lower fidelity between the two datasets. The FID between a distribution with mean $\mu_1$ and covariance $C_1$ and a distribution with mean $\mu_2$ and covariance $C_2$ is defined as follows: 
\begin{equation}
\label{eq:fid}
\text{FID}=  \norm{\mu_1-\mu_2}_2^2 + \text{tr}(C_1 + C_2 - 2\sqrt{C_1C_2})
\end{equation}

We first select the best learner by the lowest FID and add it to the ensemble. We then try adding the remaining learners to the ensemble and choose the one that most improves the FID. This process stops as soon as adding additional members does not improve FID further (Algorithm~\ref{alg:greedy}). Generally, this yields relatively small ensembles consisting of 2-3 members.

\vspace{2mm}

\noindent\textbf{Discriminator.} \sol's discriminator is a multi-layer perceptron featuring one hidden layer as consistent with previous work~\cite{silver2023mosaiq}. The input layer and hidden layer both feature the ReLU activation function, while the output layer features a sigmoid activation. The discriminator is trained on the (scaled) PCA features of the training data and the generated PCA features of the quantum generator.

\vspace{2mm}

\begin{figure}
    \centering
    \includegraphics[scale=0.31]{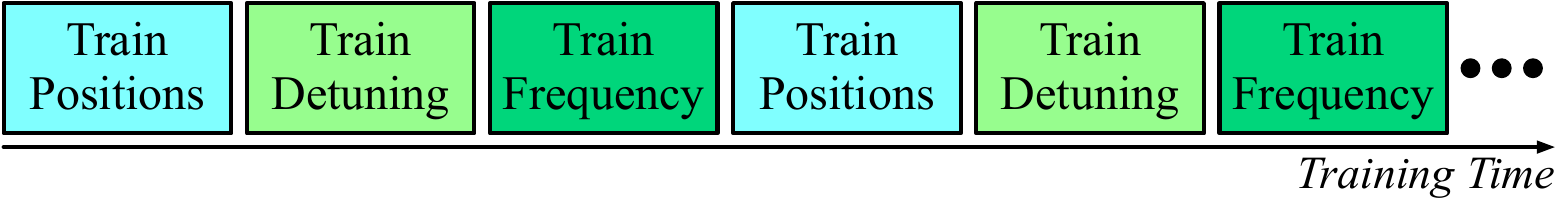}
    \vspace{0.5mm}
    \hrule
    \vspace{-3.5mm}
    \caption{\sol{} uses layered training for its parameters.}
    \label{fig:layers}
    \vspace{-4mm}
\end{figure}

\noindent\textbf{Training.} The generator and discriminator are trained in an alternating fashion, following the general outline of~\cite{goodfellow2014generative}. We use binary cross-entropy to construct the loss functions for both the generator and discriminator. When optimizing the generator in this fashion, we initially found that convergence was quite slow, which limited our ability to train a large quantity of learners. To overcome this, \sol{} utilizes a layered training scheme, shown in Fig.~\ref{fig:layers}. This allows \sol{} to focus on improving one particular aspect of the analog computation at a time. 

\vspace{2mm}

\noindent\textbf{Inference.} Once the ensemble of generators is trained and selected, we can perform inference. To do so, we input the same noise seed to each of the generators in the ensemble and then compute the average of the resulting outputs. This output is then scaled back to the PCA feature space, where we can then apply the inverse PCA transform to recover an image.

%% file: sections/methodology.tex
\section{Methodology}
\label{sec:design}

\textbf{Classical Experimental Setup.}
All training, simulation, and evaluation code is written in the Julia programming language. During the training, all quantum calculations are done by numerical simulation, assuming ideal (error-free) conditions. We use Bloqade.jl, a Julia package from QuEra, to perform a Hamiltonian simulation of neutral atoms. The generator's parameters are optimized using Nelder-Mead \cite{neldermead}, implemented in Optimization.jl. The discriminator is implemented in Flux.jl and optimized using Adam \cite{kingma2017adam}. Loss functions for both the generator and discriminator are constructed from the standard cross-entropy calculation. We restrict our parameter search space to match the physically allowed parameter space of the Aquila machine. Inference is performed in ideal simulation, noisy simulation, and on real quantum hardware.

\vspace{2mm}

\noindent\textbf{Simulation with Hardware Errors.} To simulate hardware errors, \sol{} follows a similar model to the one that is used in \cite{lu2024digitalanalog}, which is based on errors from the Aquila Rydberg computer~\cite{wurtz2023aquila}. Generally, errors are modeled as a Gaussian perturbation added to the ideal parameters. The detuning and Rabi frequency experience a stochastic shift modeled as $\Delta + N(0,0.1 \text{ MHz})$ (both local and global detuning follow this error model) and $\Omega \cdot N(1 \,,0.01 \text{ MHz})$, respectively, while the coordinates of atom $j$ are perturbed by a Gaussian process resulting in $x_j + N(0 \,,0.1 \text{ $\mu$m})$ (the same for $y_i$ as well). Here, $N(\mu,\sigma)$ is a Gaussian distribution with $\mu$ being the mean and the $\sigma$ being the standard deviation.

\vspace{2mm}

\noindent\textbf{Quantum Hardware.} For real hardware simulation, we use QuEra's Aquila device through the Amazon Braket interface \cite{gonzalez2021cloud}. At the time of writing, Aquila is readily capable of global addressing for the Rabi frequency and detuning and features an experimental implementation of locally addressable detuning. We use 1000 shots to get an empirical estimate of the probability distribution. The modulo operation and inverse PCA transform are then applied during postprocessing to convert the quantum output into images. For hardware compatibility, we must discretize the pulses so that they can be converted into an instruction set for the laser. For some pulses, like ``triangle'' and ``trapezoid'', this is not an issue, as they are already piecewise linear functions. For others, like the Gaussian pulse, this discretization will result in a piecewise linear approximation of the pulse shape, which limits our simulation accuracy.

\vspace{2mm}

\noindent\textbf{Datasets.} We evaluate \sol{} using two datasets. We use the MNIST dataset as it presents a reasonable challenge for quantum computers. MNIST contains images of handwritten digits of size 28$\times{}$28 pixels, spanning ten classes, each class corresponding to a digit~\cite{xiao2017/online}. MNIST has been widely used for NISQ-era quantum machine learning tasks~\cite{silver2023mosaiq,silver2022quilt,wang2022quantumnas, kerenidis2018quantum}. Additionally, we use Fashion-MNIST, which also contains 28$\times{}$28 images of ten classes~\cite{heusel2017gans}. While MNIST contains digits [0-9], Fashion-MNIST contains images of clothing items such as boots, sneakers, and t-shirts.

\vspace{2mm}

\noindent\textbf{Metrics.} We analyze image quality using the \textbf{Fréchet Inception Distance (FID)}~\cite{heusel2017gans}, which is a widely-used metric to assess the image quality of generative tasks~\cite{silver2023mosaiq,karras2019stylebased}. This metric is as defined in Sec.~\ref{sec:design} (Eq.~\ref{eq:fid}). We also evaluate the \textbf{Variation Across Images} to assess the diversity of the images. For a given image $G$, variance is
\begin{equation}
    \label{eq:var}
    V = \textstyle\sum_i \textstyle\sum_j (\mu_{ij} - G_{ij})^2
\end{equation}

Here, $i$ and $j$ index over the rows and columns, respectively, and $\mu_{ij}$ refers to the average of the pixel located at index $i,j$, taken over a batch of images with different input noise seeds. This gives a simple way of determining how different a given image is from the others in a given batch. This metric was also used to evaluate MosaiQ~\cite{silver2023mosaiq}. We use it to compare ideal vs. error-prone results.

\vspace{2mm}

\noindent\textbf{Competitive Techniques.} We compare against the state-of-the-art technique in quantum GANs: MosaiQ~\cite{silver2023mosaiq}. In the MosaiQ paper, it is compared against and shown to outperform prior techniques~\cite{huang2021experimental,tsang2023hybrid}, including classical and PCA inverse techniques. Therefore, we do not compare \sol{} against those techniques as MosaiQ already outperforms them. Note: MosaiQ is designed for superconducting-qubit technology and, therefore, cannot be executed using Rydberg atom technology. However, as \sol{} is the first work to implement GANs on neutral-atom technology, we compare it with a work run on superconducting technology.

%% file: sections/evaluation.tex
\section{Evaluation and Analysis}
\label{sec:eval}

In this section, we evaluate and analyze \sol{}'s performance.

\begin{figure}
    \centering
    \includegraphics[scale=0.55]{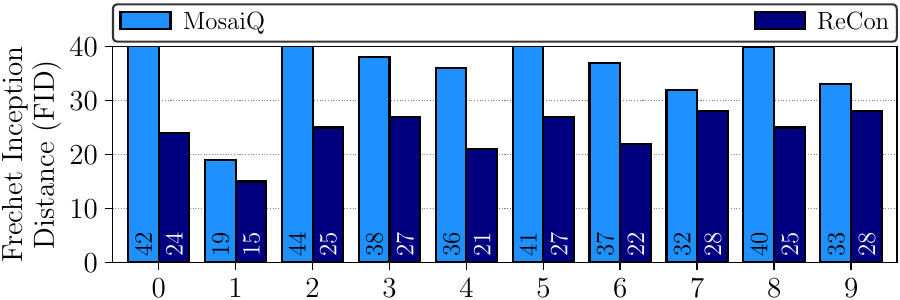}
    \vspace{0.5mm}
    \hrule
    \vspace{-3.5mm}
    \caption{\sol{} produces higher-quality images than MosaiQ across all MNIST classes despite using few PCA features.}
    \label{fig:idealmnist}
    \vspace{-4mm}
\end{figure}

\begin{figure}
    \centering
    \includegraphics[scale=0.55]{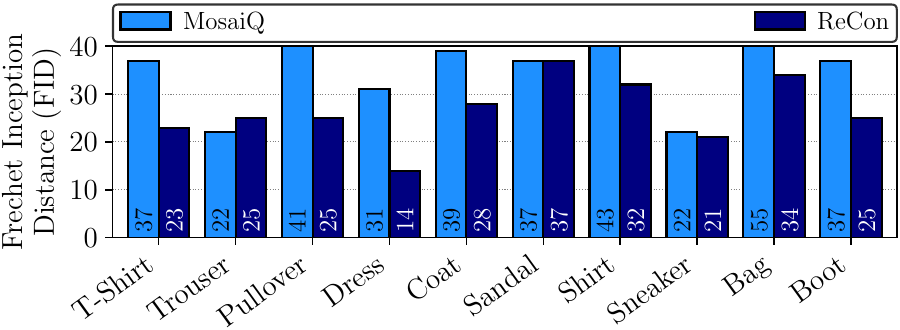}
    \vspace{0.5mm}
    \hrule
    \vspace{-3.5mm}
    \caption{\sol{} matches or beats MosaiQ across all FashionMNIST classes except for ``Trousers.''}
    \label{fig:idealfashion}
    \vspace{-4mm}
\end{figure}

\vspace{1mm}

\noindent\textbf{Ideal FID comparison with MosaiQ.} We find that \sol{} yields substantially better image quality than MosaiQ. In Fig.~\ref{fig:idealmnist}, we compare FID scores across the MNIST classes, while Fig.~\ref{fig:idealfashion} contains the same comparison for FashionMNIST. The MosaiQ scores are taken directly from the reported results in the paper~\cite{silver2023mosaiq}. Both FID comparisons assume ideal, error-free hardware. A lower FID score is better. We see \sol{} outperforms MosaiQ, having lower FID scores across all classes except for ``Trousers'' from FashionMNIST. There are fluctuations from class to class, which is attributable to the relative difficulty in generating certain classes over others. \sol{} manages to be below 30 for all MNIST classes, while MosaiQ only manages the same on class ``1''. \textit{On average, \sol{} scores 24.2 on MNIST classes, while MosaiQ scores 36.2, yielding a significant \fidpercentage{} reduction in average FID (33\%).}

\sol{} outperforms MosaiQ despite \sol{} only using $n=4$ qubits, compared to MosaiQ's $n=5$ per circuit with $8$ total circuits. Part of the reason that \sol{} does better is \sol{}'s amplitude encoding scheme, which allows it to encode more features in a given number of atoms ($2^n$, compared to MosaiQ's $n$). Even so, MosaiQ still has more than twice the number of PCA features compared to \sol{}. MosaiQ breaks up the generation task amongst several ``subgenerators'', which are separately trained to produce a subset of the PCA features. Effectively, this means that during training, MosaiQ's subgenerators learn to establish correlations between the PCA features for which they are individually responsible, but they do not strongly correlate features across different subgenerators.

\begin{figure}
    \centering
    \includegraphics[scale=0.55]
    {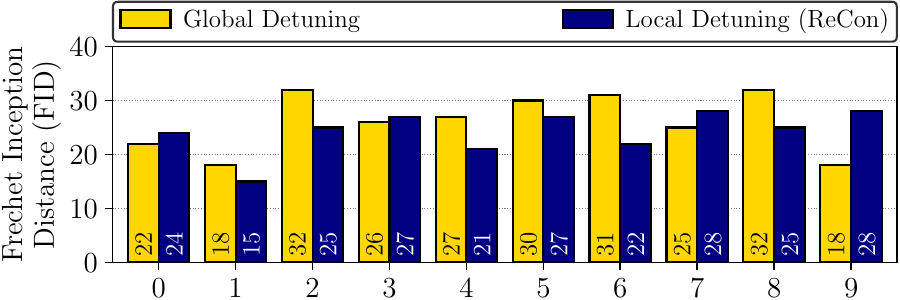}
    \vspace{0.5mm}
    \hrule
    \vspace{-3.5mm}
    \caption{A comparison of \sol{}'s usage of local detuning to a variant of our approach using global detuning.}
    \label{fig:globalvslocal}
    \vspace{-4mm}
\end{figure}

\vspace{2mm}

\noindent\textbf{Ablation: Local vs Global detuning.} We also investigate the importance of local detuning to \sol{}'s design. Fig.~\ref{fig:globalvslocal} compares local detuning (\sol{}) to a variant of our approach that uses only global detuning. In this variant, the global detuning is no longer constant in time but takes profiles similar to the pulse shapes in Fig.~\ref{fig:shapes}. Global detuning is allowed to take positive and negative values and is not required to start and end at 0, so overall, the pulses are less constrained than in the local case. Of course, the trade-off is that we lose the ability to individually address atoms, meaning all atoms would experience the same detuning. \textit{Ultimately, the two approaches are comparable, with \sol{} having an average FID of 24.2 and the global variant having an average FID of 26.1.} As improvements are made to Aquila's local detuning capabilities, we expect to see greater improvement.

\begin{figure}
    \centering
    \includegraphics[scale=0.55]{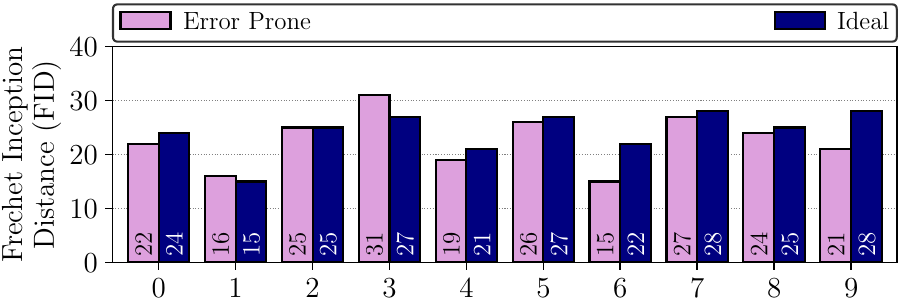}
    \vspace{0.5mm}
    \hrule
    \vspace{-3.5mm}
    \caption{\sol{}'s performance is consistent or improved under the influence of hardware errors for Fashion-MNIST.}
    \label{fig:mnistnoisy}
    \vspace{-4mm}
\end{figure}

\begin{figure}
    \centering
    \includegraphics[scale=0.55]{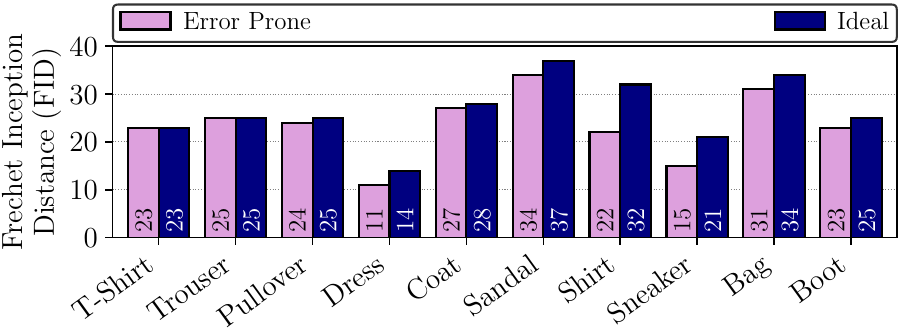}
    \vspace{0.5mm}
    \hrule
    \vspace{-3.5mm}
    \caption{Similarly, \sol{}'s performance is consistent or improved under hardware errors for Fashion-MNIST.}
    \label{fig:fashionnoisy}
    \vspace{-4mm}
\end{figure}

\vspace{2mm}

\noindent\textbf{FID for Simulated Error-Prone Hardware.} To investigate the effects of hardware noise on \sol{}, we simulated hardware errors by perturbing parameter values as described in~\cite{lu2024digitalanalog}. We do this for both MNIST (Fig.~\ref{fig:mnistnoisy}) and FashionMNIST (Fig.~\ref{fig:fashionnoisy}). Remarkably, we find that for many of the classes, this improves the final testing FID. We attribute this to an increase in image variability stemming from the randomness of the hardware errors. This randomness introduces more variety to the generated images, which in turn improves the FID score.

\begin{figure}
    \centering
    \includegraphics[scale=0.55]{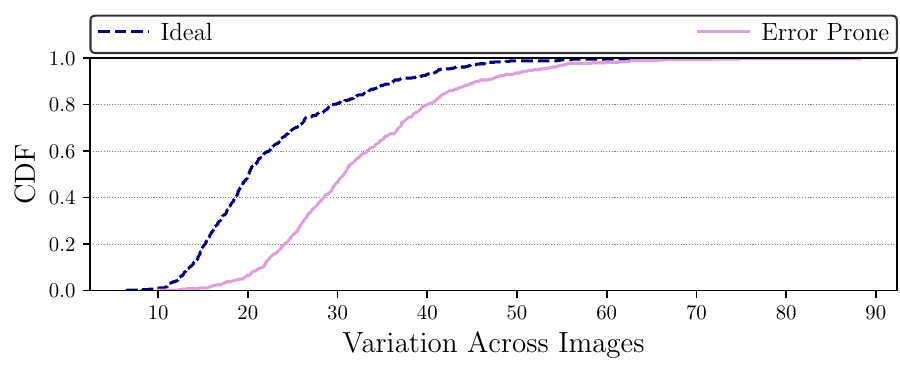}
    \vspace{0.5mm}
    \hrule
    \vspace{-3.5mm}
    \caption{Hardware errors introduce more variability into \sol{}'s generated output of MNIST class 5, increasing the diversity of the generated images.}
    \label{fig:varcdf}
    \vspace{-4mm}
\end{figure}

\begin{figure}
    \centering
    \includegraphics[scale=0.55]{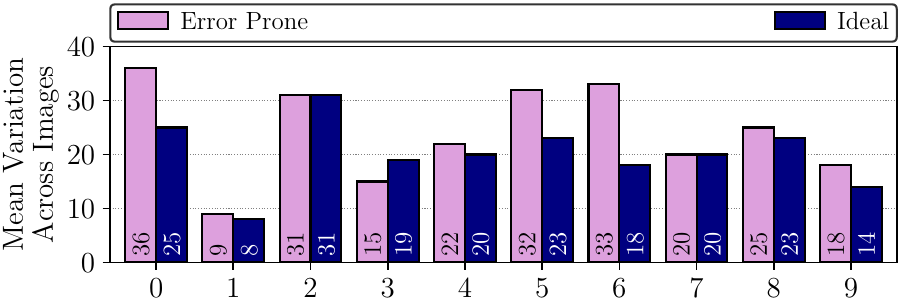}
    \vspace{0.5mm}
    \hrule
    \vspace{-3.5mm}
    \caption{Hardware errors increase \sol{}'s variability across most MNIST digit classes.}
    \label{fig:mnistvar}
    \vspace{-4mm}
\end{figure}

\begin{figure}
    \centering
    \includegraphics[scale=0.55]{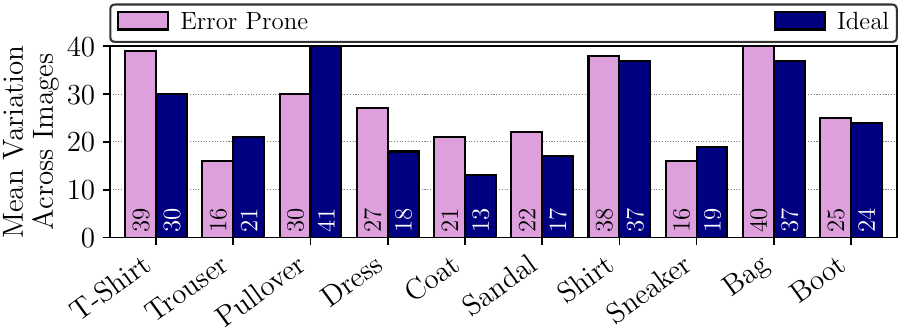}
    \vspace{0.5mm}
    \hrule
    \vspace{-3.5mm}
    \caption{Hardware errors also increase variability across many of the FashionMNIST classes.}
    \label{fig:fashionvar}
    \vspace{-4mm}
\end{figure}

\vspace{2mm}

\noindent\textbf{Variation Introduced by Hardware Error.} We can quantify the aforementioned increase in variation by considering the observed cumulative distribution function (CDF) of the variation across images. To compute this, we generate a batch of images from the generator and compute the variation of each image as defined in Eq.~\ref{eq:var}. We then construct a CDF from these scores. We plot an example of this in Fig.~\ref{fig:varcdf}, for MNIST class ``5''. In this figure, we can see that the CDF for the noisy batch rises more slowly, indicating higher overall variability in the simulation with hardware errors. This information is more compactly represented in Figs.~\ref{fig:mnistvar} and~\ref{fig:fashionvar}, which show the average variation scores for ideal and noisy simulation across both MNIST and FashionMNIST. Here, we see that hardware errors do increase the variation of generated images across most of the classes, particularly those from MNIST.

\begin{figure}
    \centering
    \includegraphics[scale=0.6]{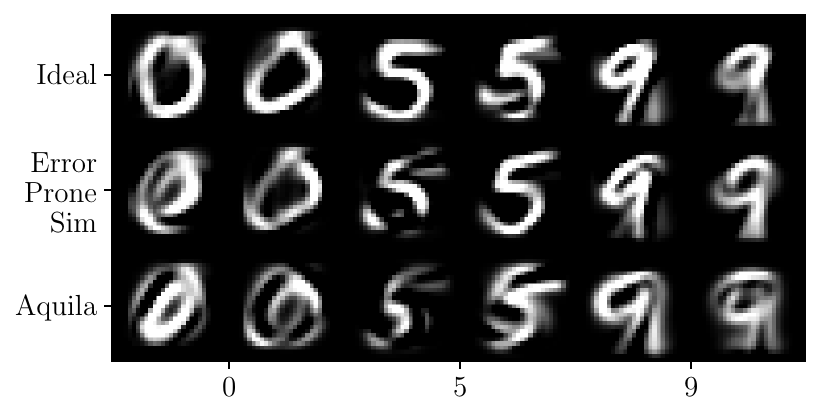}
    \vspace{-0.5mm}
    \hrule
    \vspace{-3.5mm}
    \caption{MNIST images generated by \sol{} in ideal simulation, error-prone simulation, and on real hardware (Aquila).}
    \label{fig:generatedMNIST}
    \vspace{-4mm}
\end{figure}

\begin{figure}
    \centering
    \includegraphics[scale=0.6]{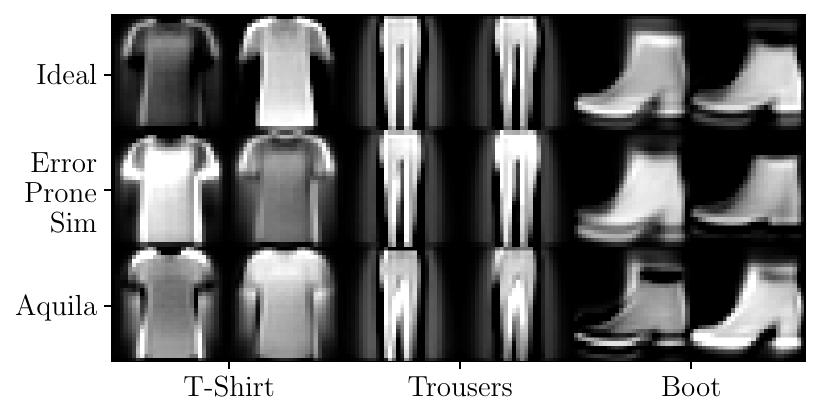}
    \vspace{-0.5mm}
    \hrule
    \vspace{-3.5mm}
    \caption{Fashion images generated by \sol{} in ideal simulation, error-prone simulation, and on real hardware (Aquila). Overall, we see that \sol{} is capable of generating diverse yet relatively high-quality images.}
    \label{fig:generatedFashion}
    \vspace{-4mm}
\end{figure}

\vspace{2mm}

\noindent\textbf{Generated Images on Real Hardware.} Sample images produced by simulation and real hardware are shown in Fig.~\ref{fig:generatedMNIST}, showing \sol{}'s capability of producing diverse images from different starting seeds. We see that while there is a degrading of quality for the MNIST images generated by Aquila, overall, they are still recognizable.
We also see relatively good performance for the FashionMNIST dataset in Fig.~\ref{fig:generatedFashion}. The figure especially demonstrates high-quality visual performance for the ``T-shirt'' and ``Boot'' classes due to their recognizable shapes when generated on Aquila.

%% file: sections/related_work.tex
\section{Related Work}
\label{sec:design}

\textbf{Machine Learning on Rydberg Atom Quantum Computers.}  A recent work~\cite{lu2024digitalanalog} explored implementing hybrid digital-analog algorithms on Rydberg atom platforms for machine learning tasks. This digital-analog scheme utilizes alternating layers of tunable digital and analog computing to create a learning circuit, which is then optimized similarly to a VQA. Using both ideal and noisy simulation, this method was then benchmarked on multiple tasks, including a simplified version of MNIST classification. While hybrid digital-analog approaches are exciting contenders for near-term hardware, current commercially available Rydberg atom computers like Aquila do not yet have the capability to operate in the digital mode, thus preventing hybrid schemes from being implemented.

\vspace{2mm}

\noindent\textbf{Generative Adversarial Networks on Quantum Computers.}
In addition to MosaiQ~\cite{silver2023mosaiq}, there have been several other implementations of quantum GANs. Huang et al.~\cite{huang2021experimental} experimentally implemented a GAN using superconducting qubits. Both the generator and discriminator are real quantum circuits. The generator is trained using quantum gradients to replicate the classical XOR function. Hu et al.\cite{hu2019quantum} also implement a GAN on superconducting hardware. They focus on learning to generate mixtures of quantum states. Tsang et al.~\cite{tsang2023hybrid} numerically explore a quantum GAN architecture that breaks images up into patches, which are trained on separate subgenerators. They test this design on both MNIST and FashionMNIST. Their design also allows for multiple classes to be produced by the same generator. Unlike \sol{}, these works are not designed for Rydberg atom quantum computing.


%% file: sections/conclusion.tex
\section{Conclusion}
\label{sec:design}

In this paper, we proposed \sol{}, the first work to develop and demonstrate GANs for analog Rydberg atom quantum computers. \sol{} implements optimizations such as noise injection at pulse starting points, parameterization of spatial and temporal features, and ensemble training and selection for improved quality and variety in generated images. \sol{} is able to achieve \fidpercentage{} lower FID than the state-of-the-art technique implemented on superconducting-qubit quantum computers for the MNIST dataset. We hope our work will set the stage for using Rydberg atom quantum computers for generative and other machine-learning tasks.

\section*{Acknowledgement}

This work was supported by Rice University. We would like to thank the anonymous reviewers for their insightful feedback. We would also like to thank the AWS Braket Cloud team for their support in running our workloads on the QuEra Aquila quantum computer. The views expressed are those of the authors and do not reflect the official policy or position of the AWS or QuEra teams.